\documentclass[sigconf]{acmart}
\AtBeginDocument{%
  }

\setcopyright{none}

\setcopyright{acmlicensed}
\copyrightyear{2026}
\acmYear{2026}
\acmConference[AutomotiveUI '26 AnthroAssist Workshop: Anthropomorphic and Intelligent Assistive Systems for Automated Vehicles]{}{Sept 20, 2026}{Gothenburg, Sweden}

\begin{document}

\title[]{The Agent Is Not Your Driver: Clarifying the Implied Capabilities of In-Vehicle Anthropomorphic Agents}
\thispagestyle{empty}

\author{Benjamin Hardin}
\email{benjamin.hardin@cs.ox.ac.uk}
\orcid{0009-0000-9938-3336}
\correspondingauthor
\affiliation{%
  \institution{Department of Computer Science,}
  \institution{University of Oxford}
  \city{Oxford}
  \country{UK}
}


\begin{abstract}
  This position paper considers how the gap between an anthropomorphic agent in an autonomous vehicle and the automated driving system of the vehicle can create psychological risks if the agent inaccurately presents itself as integrated into the driving system. This gap is under-explored both because of the evolving nature of these agents and because of study limitations in Human-Computer Interaction (HCI) research. Nevertheless, this gap may cause the agent to misrepresent itself as more knowledgeable about the car than it is, creating a risk of dependence and, in the worst case, betrayal if the occupant feels that the agent has misled them. This paper proposes a research agenda for future work to understand the psychological risk that exists in the gap between agents and driving systems and to narrow this gap in a manner aligned with safety assurance. As these agents evolve to hold greater capabilities and increased anthropomorphism, future research must investigate risks which may emerge from richer interaction with them. The impact of how an agent presents itself and its knowledge becomes greater.
\end{abstract}




\maketitle

\section{Introduction}

The integration of conversational agents into autonomous vehicles (AVs) is expected to provide high levels of user experience benefits, albeit with considerations for how to make the agents respectful, friendly, and useful, particularly across cultures that may value agents differently \cite{SOGEMEIER2025279, large2019please, examining_cross, anthropomorphic_intelligent}. The discussion around agents has also focused primarily on the truthfulness of agents, ensuring that agents provide honest and clear communication about the vehicle to occupants. However, in the discussion around considerate and truthful communication, there remains a large gap: an inherent deception in the reality that these agents are not part of the driving system itself.

Fundamentally, there is a gap between the agent and the autonomous functions of the AV which may lead to errors in agent communication. The agent is not integrated into the core driving system of the vehicle, and the agent is not driving the vehicle; 
the agent sits as a separate entity that reasons about the driving system while interacting with occupants. Despite this large disconnect, it has not been studied how occupants of a vehicle react to understanding that this disconnect exists. This disconnect is not communicated to occupants, and it has not been studied whether they naturally discover this gap through some behaviour or limitation of the agent. 

Furthermore, these agents are not expected to be conversational alone, but are expected to take on some personified form, either virtual or physically embodied in the vehicle. Adding anthropomorphic features has already been shown in studies to affect perception of the voice agent in the vehicle compared to no anthropomorphism, inspiring trust and more positive attitudes as well as increasing estimated intelligence of the AV \cite{chu2024intersection, ban2025vehicle}. Anthropomorphic assistants are no longer only in research settings, as companies such as NIO, Mercedes-Benz, and MINI already offer virtual characters in their vehicle infotainment systems: the NOMI, MBUX Virtual Assistant, and Spike, respectively \cite{nio_nomi_2020, mercedes_benz_virtual_assistant_2024, kremser_mini_spike_2023}. Admittedly, these agents are relegated mostly to infotainment functions, rather than driving functions, although one can expect their usage to be expanded in the driving direction in the near future as we have seen in research settings already \cite{lawson2023anthropomorphic, lawson2023embodying}.

This position paper considers this gap in capabilities and functions of the in-vehicle agents and the autonomous driving system itself. Throughout this piece, an embodied in-vehicle agent will be referred to as an \textbf{anthropomorphic agent (AA)}, and these agents will be defined by three primary attributes: 1) their location in the vehicle cabin, 2) their appearance as an avatar, whether in an infotainment screen or in another part of the vehicle, and 3) their ability for two-way communication with the user about driving functions of the vehicle, among other topics. The driving functions of vehicles will be referred to as the \textbf{automated driving system (ADS)} throughout this piece, and the ADS is defined as all the various intelligence models and systems that the vehicle uses to perform autonomous driving functions, whether partial or full autonomy. This paper focuses on in-vehicle AAs, although AAs on external vehicle interfaces certainly represent an important area of research for future discussion.

\section{The Gap in Anthropomorphic Agents' Understanding and Confidence}
The first fundamental concern about anthropomorphic agents in vehicles which this paper presents is that there is a gap between the AA and the ADS which may lead to inaccuracies and errors. In the vast majority of literature around AAs thus far, the AA is a separate entity from the ADS, albeit with varying degrees of communication with the ADS. In the base case, the AA is completely separate from the ADS, and the only questions it can answer about the ADS and its decisions are based solely on observations of behaviours of the ADS. This has been seen already in production systems such as the Wayve LINGO-1 system, which offers black-box explanations, although not in an anthropomorphic form \cite{wayve2023lingo}. This is most similar to if another passenger were to comment on the vehicle, although the distinction is that the AA comments with authority as part of the vehicle system and declares a level of understanding of vehicle decision making. The next level of integration is when an AA has access to some explanations generated by the driving system itself, such as attention maps from a CNN, and uses these explanations to form its response. To date, this hybrid approach has not been demonstrated for an AA, only for textual or audible explanations. Finally, one can imagine an AA that is directly a component of the core model of the ADS, able to influence decisions made by the ADS about how to drive and behave. Some have already proposed embedding language models into the ADS, and one can imagine these models being exposed as an AA to the occupant \cite{mao2023language, mao2023gpt, sima2024drivelm}.

However, despite this separation between the AA and ADS, many AAs are intentionally designed to appear as a direct anthropomorphic representation of the ADS. They present as an embodiment of the vehicle and can be queried about the vehicle. One might ask what the vehicle plans to do, why it took certain actions, or how the vehicle functions. The AA may answer these questions without access to the real answers, which lie in the ADS, a separate component. 
One work sought to separate the agent into two agents, one for the ADS and one for the infotainment, but found that this approach produced no significant differences in trust \cite{torggler2022beyond}. Although the experiment was designed with non-anthropomorphic agents (i.e. conversational only), the result that trust did not change between the two configurations highlights that separating the AA from the ADS likely will not work alone. Alternatively, if the AA is separate from the ADS, then it should provide a more honest description of its abilities and limitations.

Finally, these agents lack a true understanding of ADS confidence ratings and the accuracy of explanations, simply because they observe the ADS from an outside, or black-box, perspective. This remains, and will likely continue to be, an open problem because it is dependent on the chosen architecture of the ADS. As architectures change and evolve over time, so does their interpretability. However, what can and should be studied is how agents can express their confidence and understanding to occupants of the vehicle. It will be essential to understand what conversational patterns are necessary to ensure occupants have an accurate understanding of the ADS and do not ascribe too much authority to the AA. This psychological component will now be discussed in greater detail.

\section{Psychological Safety Effects of Using Anthropomorphic Agents Should Not Be Ignored}
The second fundamental concern about anthropomorphic agents in vehicles which this paper presents is the psychological safety of using these agents. Psychological safety has been considered in teamwork settings \cite{edmondson1999psychological}, robotic applications \cite{lasota2017survey, rubagotti2022perceived, kamide2012new}, and autonomous driving contexts \cite{sirgabsou2025av} as a way to consider the various risks that can arise from psychological interactions. Many of these risks are related to trust in other agents, whether human or robotic, to perform as safely as expected. Trust has been studied extensively for AVs and for AAs in AVs, but psychological safety goes beyond trust or comfort alone to include aspects such as risk arising from deception or surprise caused by unpredictable actions \cite{sirgabsou2025av}. These aspects may not be inherently related to trust but can still pose psychological risks such as surprise from unpredictable communication or betrayal from misrepresentation. Misrepresentation is perhaps a much greater psychological safety risk than merely uncomfortable communication, as it may represent the AA acting as if it has control over and understanding of the driving when it actually has neither. This risk may be exacerbated by the fact that AAs have been shown to increase over-reliance by drivers on ADS features \cite{ban2025vehicle}.

Psychological safety during AA–human interaction does not appear to have been sufficiently examined so far. At a basic level, the most important components of psychological safety relevant to in-vehicle anthropomorphism include deception, comfort, trust, and blame. In-vehicle AAs resemble chauffeurs in some sense as they help the occupant navigate and drive to a destination and engage in two-way communication to refine and reach this goal. 
Therefore, in the same manner as interpersonal psychology, individuals may use communication to deceive one another, either intentionally or unintentionally, or may not make the other feel safe to express their needs. Furthermore, human-like AAs have been shown to increase trust compared to other AAs \cite{lawson2023embodying, liu2025better}. This increased trust combined with closer-to-human-like relationships raises risks of intense yet faulty communication and, in the worst case, AI psychosis (which has already occurred with conversational agents in other domains \cite{hudon2025delusional, flathers2026beyond}).

Approaching concerns of AA communication and presentation through the lens of psychological safety is a necessary aspect of future research for three reasons. First, considering psychological safety helps researchers consider psychological components that they would not normally have considered. It can be natural for one to focus solely on trust and miss other components such as comfort or predictability, which, if neglected, can lead to a negative user experience. Second, considering psychological safety helps analyse the components not only in isolation, but also in relation to one another, such as how in-depth explanations can increase trust at the expense of mental workload. Psychological safety can highlight potential trade-offs in prioritising one component over another. Many of these trade-offs are not currently recognised or measured, and there is a strong opportunity for future research in this area. Third, considering psychological safety allows for a safety assurance engineering approach to these issues. When one considers a single component on its own, it is easy to view it as a scale from bad to good or low to high. But what does low trust or low comfort mean practically for safety assurance of a system? Using the concepts of psychological safety allows one to introduce the concept of psychological hazards and risks in order to scope how low comfort, for instance, can create tangible safety risks rather than abstract measures. Safety assurance methods such as System-Theoretic Process Analysis (STPA) and Safety of the Intended Functionality (SOTIF) can be used when one views these concerns as psychological risks \cite{sirgabsou2025av}. With regard to anthropomorphism specifically, a safety assurance approach allows consideration of how this concern over agent deception creates psychological risks which can be categorised by severity and tackled systematically to avoid psychological harm. Furthermore, integration with SOTIF allows for consideration of how psychological factors may affect physical safety, such as how shock or surprise may affect reaction times and takeover quality of drivers.


\section{A Proposed Research Agenda}
\subsection{Measuring Psychological Safety of Agent Interaction}
It is important to verify how substantial this psychological safety risk is before considering methods for mitigating it. Fortunately, variations of standard user studies in AV simulators or test vehicles represent a viable means of assessing this risk. Imagine a future study that considers how participants’ knowledge of the level of disconnect between an agent and the vehicle driving algorithm influences their psychological safety. Ideally, this study should be run with multiple styles of agents, some of which have no connection to the ADS, some of which are slightly embedded in the ADS, and some of which are completely embedded. For a study in the short term, all of these can be Wizard-of-Oz if they appear to function correctly. For each type of agent, some participants (the control group) should not be told about the level of connection between the AA and the ADS. 
Other participants in the experimental group should be told about the level of connection between the AA and the ADS during the drive. If enough participants are recruited, different styles of communicating the disconnect may be tested, such as a simple statement versus a simple technical chart of AA–ADS interaction. Participants' psychological safety should be measured before and after they are informed of this separation between the AA and the ADS to understand how their trust and risk perception change.

While there are generally no universally accepted measurements for in-vehicle psychological safety, existing scales related to components of psychological safety can be used. These include scales for trust (such as the Situational Trust Scale for Automated Driving (STS-AD) \cite{sts-ad}, Trust in Automated Driving (TiAD) Scale \cite{tiad}, or Trust of Automated Systems Test (TOAST) \cite{toast}), perceived risk scales \cite{li2019no}, and control and transparency scales \cite{schneider2023don}. Finally, safety engineers of these systems can use safety assurance frameworks to measure psychological risk \cite{sirgabsou2025av}.

\subsection{Agent Connection to the Driving Algorithms}
Concerning technical improvements to AAs, research must consider how these agents can safely be connected to the driving algorithms themselves, either through more standardised, informed, and robust communication channels or through direct embedding into the ADS. Directly embedding an agent into the ADS carries some risks, despite being the most likely way to accurately represent the ADS behaviour. Namely, if communication between the ADS and the AA is two-way, then the occupant may be able to influence vehicle decision making intentionally or unintentionally merely through communication with the AA. While having an AA embedded into the ADS might be beneficial in certain situations, such as a passenger specifying where in front of a building they would like to be dropped off, it carries risks in other situations such as if the passenger were to complain about how carefully the vehicle was driving and the agent used this as a reason to drive more aggressively, cutting off other traffic or taking unsafe actions. For this reason, many studies have not directly connected AAs to the ADS, although many ADSs have been shown to be able to function with inputs from LLMs or language agents, the difference being that these agents do not interact with the passenger \cite{wayve2023lingo, sima2024drivelm, mao2023gpt, mao2023language}. Simply, allowing any form of conversational agent a level of control over the ADS is likely to pose immense safety assurance challenges related to verification of commands from the occupant.



\subsection{Communication of Limitations and Understanding}

Although general communication strategies and preferences of AAs are actively being studied \cite{examining_cross}, the design of AAs to explain ADS decision making along with the understandability of this communication is less studied. As a second research agenda point, it will be important to understand how well the gap between the AA and ADS is currently communicated and understood by vehicle occupants. Communication strategies should be tested with participants across cultures, as various cultures exhibit strong differences in AA preferences as was previously referenced. As it is now widely accepted by the general public that LLMs hallucinate, can we expect the same for in-vehicle agents? Will the public accept that in-vehicle agents do not accurately portray the thinking and actions of the vehicle, or will this be deemed unacceptable because it is a safety-critical domain? Furthermore, will occupants accept that this disconnect between the agent and the driving algorithm exists, or will it destroy all illusion of the agent as helpful? As a by-product of this improved channel, might the AA make the occupant more receptive to ADS information?

From a communications and reliance standpoint, this implication that the AA is directly connected to the ADS might be seen as similar to deceit or betrayal from someone claiming they are responsible for work they did not do. In interpersonal contexts, this has been shown to lead to a loss of trust and undermine goals \cite{borthakur2026mindfulness}; in the worst case it may lead to dissonance and moral injury \cite{layson2026unexpected}. It will be important to understand whether these same stakes apply to a highly conversational AA in an AV as they have been shown to apply to human interpersonal contexts.




\section{Concluding Remarks and Future Directions}
In-vehicle anthropomorphic agents (AAs) are a novel and rapidly expanding technology that offers many potential benefits to the in-vehicle experience. They may help alert occupants to dangerous situations, provide an easy, natural-language way to understand and customise the ride, or reduce workload in handover and difficult situations, all while improving trust, enjoyment, and sense of control \cite{large2019please, lawson2023anthropomorphic}. However, this paper has argued that the inherent disconnect that exists between these agents and the vehicle itself creates a divide of deception that poses a psychological risk to occupants. This psychological risk has been under-explored thus far, and Wizard-of-Oz simulator studies have not allowed researchers to accurately draw conclusions about this risk. It is important for researchers to explore this divide and tackle it before these agents become a standard element of the AV experience. To this end, this paper has proposed a basic research agenda which I hope will guide these agents away from posing psychological risk. This paper has considered only in-vehicle agents, but agents on external human-machine interfaces (eHMIs) will also require consideration of psychological risks to other road users which may impact other parts of the field. An era of anthropomorphic and conversational autonomous vehicles is well under way, and as researchers, we have a responsibility to ensure that these new and exciting forms of agents introduce as little risk as possible. Our designs should be transparent without overwhelming the user, and in this regard, there is work to be done for anthropomorphic agents in vehicles. 

\bibliographystyle{ACM-Reference-Format}
\bibliography{sources}










\end{document}